\documentclass[12pt]{article}
\usepackage{epsfig}
\usepackage{hyperref}
\usepackage{amssymb}
\usepackage{amsmath}
\usepackage{amsfonts}
\usepackage{latexsym}
\usepackage{wasysym}
\usepackage{multirow}
\usepackage{fixmath}
\usepackage{color}
\usepackage{stackrel}
\usepackage{txfonts}                                                            
\usepackage{bbm}
\usepackage{mathrsfs}
\newcommand{\bea}{\begin{eqnarray}}
\newcommand{\eea}{\end{eqnarray}}
\newcommand{\bite}{\begin{itemize}}
\newcommand{\eite}{\end{itemize}}

\newcommand{\msb}{\overline{\rm MS}}

\newcommand{\kappas}{\kappa_{\rm sea}}

\newcommand{\lambdas}{\lambda_{\rm sea}}
\newcommand{\lambdav}{\lambda_{\rm val}}
\newcommand{\deltas}{\delta_{\rm sea}}
\newcommand{\deltav}{\delta_{\rm val}}

\begin{document}

\title{
\vspace{-2.5cm} 
\flushleft{\normalsize ADP-14-27/T885} \\
\vspace{-0.35cm}
{\normalsize DESY 14-172} \\
\vspace{-0.35cm}
{\normalsize Edinburgh 2014/14} \\
\vspace{-0.35cm}
{\normalsize Liverpool LTH 1024} \\
\vspace{-0.35cm}
{\normalsize October 2014} \\
\vspace{0.75cm}
\centering{\Large \bf A novel approach to nonperturbative renormalization of singlet and nonsinglet 
lattice operators}\\[1em]} 

\author
{A.~J.~Chambers\\
\small
{CSSM, Department of Physics, University of Adelaide, Adelaide SA 5005, Australia} \\
\and R.~Horsley  \\
\small
{School of Physics and Astronomy, University of Edinburgh, Edinburgh EH9 3JZ, UK} \\
\and Y.~Nakamura \\
\small
{RIKEN Advanced Institute for Computational Science, Kobe, Hyogo 650-0047, Japan}\\
\and
H.~Perlt\\
\small
{Institut f\"ur Theoretische Physik, Universit\"at Leipzig, 04103 Leipzig, Germany}
\and
P.~E.~L.~Rakow  \\
\small
{Theoretical Physics Division, Department of Mathematical Sciences, }\\
\small
{University of Liverpool, Liverpool L69 3BX, UK}\\
\and
G.~Schierholz  \\
\small
{Deutsches Elektronen-Synchrotron  DESY, 22603 Hamburg, Germany }
\and
A.~Schiller  \\
\small
{Institut f\"ur Theoretische Physik, Universit\"at Leipzig, 04103 Leipzig, Germany }
\and
J.~M.~Zanotti  \\
\small
{CSSM, Department of Physics, University of Adelaide, Adelaide SA 5005, Australia}
}

\date{}

\maketitle

\vspace*{0.5cm}

\begin{center}
{\large QCDSF Collaboration}
\end{center}

\clearpage
\begin{abstract}
A novel method for nonperturbative renormalization of lattice operators is introduced, 
which lends itself to the calculation of renormalization factors for nonsinglet as well 
as singlet operators. The method is based on the Feynman-Hellmann relation, and 
 involves computing two-point correlators in the presence of generalized 
background fields arising from introducing additional operators into the action. 
As a first application, and test of the method, we compute the renormalization 
factors of the axial vector current $A_\mu$ and the scalar density $S$ for both 
nonsinglet and singlet operators for $N_f=3$ flavors of SLiNC fermions. For nonsinglet operators, 
where a meaningful comparison is possible, perfect agreement with recent calculations using 
standard three-point function techniques is found.
\end{abstract}

\clearpage
\section{Introduction}

To relate bare lattice results of hadron matrix elements and decay constants to phenomenological 
numbers, which are usually given in the $\msb$ scheme, the underlying operators need to be renormalized. 
This requires a nonperturbative method, because lattice perturbation theory is considered to be 
unreliable at present couplings.

A general nonperturbative method is the RI$^\prime$-MOM subtraction scheme, which has been 
proposed in~\cite{Martinelli:1994ty}, with some refinements being added in~\cite{Gockeler:1998ye}. 
Starting from the bare vertex function
\begin{equation}
\Gamma_{\mathcal{O}}(p) = S^{-1}(p)\,G_{\mathcal{O}}(p)\,S^{-1}(p) \,,
\label{gamma}
\end{equation}
where
\begin{equation}
G_{\mathcal{O}}\,(p) = \frac{1}{V}\,\sum_{x,y,z}\, e^{-ip(x-y)} \langle q(x)\, \mathcal{O}(z)\, \bar{q}(y) \rangle 
\end{equation}
is the quark Green function with operator insertion $\mathcal{O}$, and
\begin{equation}
S(p) = \frac{1}{V}\,\sum_{x,y}\, e^{-ip(x-y)} \langle q(x)\, \bar{q}(y) \rangle 
\end{equation}
is the quark propagator, the renormalized vertex function is defined by
\begin{equation}
\Gamma_{\mathcal{O}}^R\,(p) = Z_q^{-1}\, Z_{\mathcal{O}}\,\Gamma_{\mathcal{O}}\,(p) \,.
\end{equation}
$Z_q$ denotes the quark field renormalization constant, which is taken as
\begin{equation}
Z_q(p) = \frac{{\rm Tr}\left[-i \sum_\lambda \gamma_\lambda \sin (p_\lambda)\, S^{-1}(p)\right]} {12\sum_\rho \sin^2(p_\rho)} \,.
\label{eq:RIMOM2}
\end{equation}
The renormalization factor $Z_{\mathcal{O}}(\mu)$ is determined by imposing the renormalization condition
\begin{equation}
\frac{1}{12}\,{\rm Tr}\,\left[\Gamma_{\mathcal{O}}^R\,(p)\,{\Gamma_{\mathcal{O}}^{\rm Born}(p)}^{-1}\right] = 1 
\end{equation}
at the scale $p^2=\mu^2$. 
Thus
\begin{equation}
Z^{-1}_{\mathcal{O}}\,(\mu)=\frac{1}{12}\,{\rm Tr}\,
\left[\Gamma_{\mathcal{O}}\,(\mu)\,{\Gamma_{\mathcal{O}}^{\rm Born}(\mu)}^{-1}\right]\, Z^{-1}_q(\mu) \,.
\label{eq:RIMOM1}
\end{equation}
The lattice spacing $a$ is assumed to be one, if not stated otherwise. $V$ is the lattice volume. 

The evaluation of $Z_{\mathcal{O}}$ requires the calculation of three-point functions. 
In the case of flavor singlet matrix elements this entails the computation of quark-line 
disconnected diagrams, which requires inversions of the fermion matrix at every lattice 
point and still leads to a poor signal to noise ratio. In this paper we propose an 
alternative method, based on the Feynman-Hellmann (FH) relation, which eliminates the 
issue of computing disconnected contributions directly at the expense of requiring the generation of 
additional ensembles of gauge field configurations. This essentially involves computing 
two-point correlators only in the presence of generalized background fields, 
which we show arise from introducing the operator $\mathcal{O}$ into the action,
\begin{equation}
\mathcal{S} \rightarrow \mathcal{S}(\lambda) = \mathcal{S} - \lambda\,\sum_x\,\mathcal{O}(x) \,, 
\quad \mathcal{S} = \mathcal{S}_F + \mathcal{S}_G \,,
\label{action}
\end{equation}
where $\mathcal{S}_F$ and $\mathcal{S}_G$ are the fermionic and gauge field actions.
A further advantage of this method is that the signal to noise ratio will be directly proportional 
to the external parameter $\lambda$, and thus can be controlled from the outside, as opposed to the standard three-point function 
calculation. 

The quark propagators in (\ref{gamma}) are calculated by inverting the fermion matrix, 
and so must be modified if we change the quark action. This change is straightforward to apply, 
only requiring a redefinition of the Dirac operator.
In addition, any modification we make to the action in (\ref{action}) should be included during 
the generation of the background gauge fields. By choosing to neglect either one of these 
modifications, we are able to individually isolate connected and disconnected contributions 
to the vertex function. Thus, modifications to the gauge configurations allow access to 
disconnected quantities, and modifications to the calculation of propagators allow access 
to connected quantities. 

This paper follows previous work on hyperon sigma terms~\cite{Horsley:2011wr}, the glue in 
the nucleon~\cite{Horsley:2012pz}, and the spin structure of hadrons~\cite{Chambers:2014qaa}, 
already showing the potential of the Feynman-Hellmann approach to the calculation of hadron 
matrix elements. The outline of the paper is as follows. Section~\ref{sec:renorm} describes  
the Feynman-Hellmann relation as relevant for the calculation of renormalization factors. 
In Secs.~\ref{sec:A3} and \ref{sec:S} we apply the method to the computation of renormalization 
factors of the axial vector current $A_\mu$ and the scalar density $S$, respectively, 
for singlet and nonsinglet operators. The calculations are done with $N_f=3$ flavors of 
SLiNC fermions~\cite{Cundy:2009yy,Bietenholz:2011qq}. Section~\ref{sec:con} contains
our conclusions.

\section{The Feynman-Hellmann method}
\label{sec:renorm}

Throughout this paper we will consider quark-bilinear, flavor diagonal operators
\begin{equation}
\mathcal{O}(x) = \bar{q}(x)\,\Gamma\,q(x) 
\end{equation}
only, where $\Gamma$ is some combination of gamma matrices. The generalization to operators 
including covariant derivatives is straightforward. The modified fermionic action then reads
\begin{equation}
\mathcal{S}_F(\lambda)=\sum_{q=u,d,s} \sum_{x} \bar{q}(x) \left[D + M - \lambda\,\Gamma\right] q(x) \,,
\end{equation}
where $D$ is the lattice Dirac operator including the Wilson and clover terms, and $M$ is the 
Wilson mass term. The latter is a diagonal $3\times 3$ matrix in flavor space,
\begin{equation}
M = \left(\begin{tabular}{ccc} $1/2\kappa_u$ & & \\ & $1/2\kappa_d$ & \\ & & $1/2\kappa_s$
\end{tabular}\right) \,.
\end{equation}
One is mainly interested in renormalization factors in a mass-independent scheme, such as 
the $\msb$ scheme. To comply with that, we choose the quarks to be mass degenerate,
\begin{equation}
M = (1/2\kappa)\,\mathbbm{1} \,, \quad \kappa_u=\kappa_d=\kappa_s\equiv \kappa \,,
\label{mdeg}
\end{equation}
and tune $\kappa$ to its critical value, $\kappa_c$, at the end of the calculation. 
A better choice might be to only take the $u$ and $d$ quarks as mass-degenerate, 
$\kappa_u =\kappa_d \equiv \kappa_\ell$, and keep the sum of the quark masses 
fixed~\cite{Bietenholz:2011qq}, $2/\kappa_\ell+1/\kappa_s = \mbox{constant}$, 
while taking $\kappa_\ell$ to its critical value, $\kappa_{\ell,c}$. In that case we would have
\begin{equation}
M = \left(\begin{tabular}{ccc} $1/2\kappa_\ell$ & & \\ & $1/2\kappa_\ell$ & \\ & & $1/2\kappa_s$
\end{tabular}\right) \,.
\label{mmdeg}
\end{equation}

\begin{figure}[!t]
  \begin{center}        
  \includegraphics[scale=0.63,clip=]{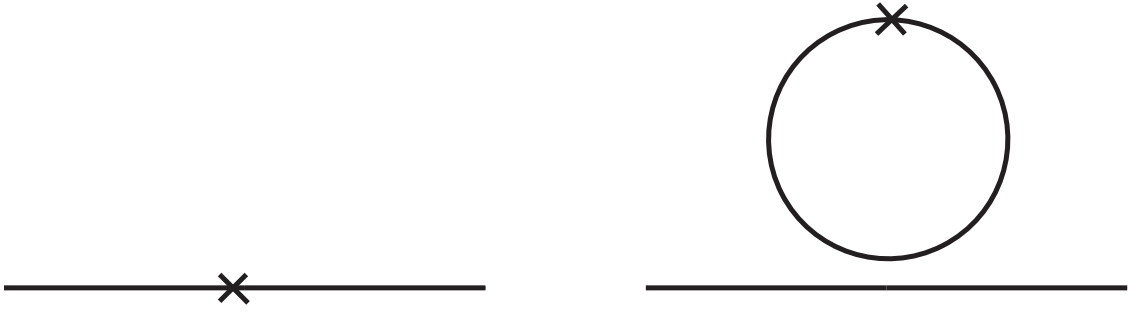}
  \end{center}
  \caption{Diagrams contributing to the renormalization of quark-bilinear operators 
  (inserted at point $\mathbf{\times}$). The left figure shows the connected (nonsinglet) 
  contribution, the right figure the disconnected (singlet minus nonsinglet) contribution. 
  Gluon lines have been omitted.}
  \label{fig:green}
\end{figure}

After integrating out the quark fields, the fermion propagator becomes
\begin{equation}
S(\lambdas,\lambdav) = \frac{\int \mathcal{D}U\,\left[D+M-\lambdav\,\Gamma\right]^{-1}\, 
\det\left[D+M-\lambdas\,\Gamma\right]\,\exp\{-\mathcal{S}_G(U)\}}
{\int \mathcal{D}U\, \det\left[D+M-\lambdas\,\Gamma\right]\, \exp\{-\mathcal{S}_G(U)\}}\,,
\end{equation}
where we differentiate between operator insertions in the quark propagator ($\lambdav$) and 
the fermion determinant ($\lambdas$), to separate connected and disconnected diagrams eventually. 
In what follows Fourier transformation of $S(\lambdas,\lambdav)$ to momentum space is understood. 
For the sake of simplicity any dependence on external momenta will be omitted. Expanding the 
propagator in terms of $\lambdas, \lambdav$ gives
\begin{eqnarray}
\begin{split}
S(\lambdas,\lambdav) &= \langle \left[D+M\right]^{-1}\rangle +\lambdav\, 
\langle\left[D+M\right]^{-1}\Gamma\, \left[D+M\right]^{-1}\rangle\\
&-\lambdas\,\left\{\langle \left[D+M\right]^{-1} {\rm Tr}\, (\Gamma\left[D+M\right]^{-1})\rangle - 
\langle \left[D+M\right]^{-1}\rangle\, \langle {\rm Tr}\, (\Gamma\left[D+M\right]^{-1})\rangle\right\} \\
&+ O(\lambdas^2,\lambdas\,\lambdav,\lambdav^2) \,,
\end{split}
\label{exp}
\end{eqnarray}
where the expectation values $\langle \cdots \rangle$ refer to the unmodified action. 
By differentiating the quark propagator with respect to $\lambdav$ and $\lambdas$ we obtain
\begin{equation}
\left.\frac{\partial\,S(0,\lambdav)}{\partial\,\lambdav}\right|_{\lambdav=0} = 
\langle\left[D+M\right]^{-1}\Gamma\, \left[D+M\right]^{-1}\rangle\, \equiv\, G_{\mathcal{O}}^{\rm con}
\end{equation}
and 
\begin{equation}
\begin{split}
\left.\frac{\partial\,S(\lambdas,0)}{\partial\,\lambdas}\right|_{\lambdas=0} = 
&-\langle \left[D+M\right]^{-1} {\rm Tr}\, (\Gamma\left[D+M\right]^{-1})\rangle\\
&+ \langle \left[D+M\right]^{-1}\rangle\, 
\langle {\rm Tr}\, (\Gamma\left[D+M\right]^{-1})\rangle\, \equiv\, G_{\mathcal{O}}^{\rm dis} \,,
\end{split}
\label{Gdisc}
\end{equation}
where $G_{\mathcal{O}}^{\rm con}$ and $G_{\mathcal{O}}^{\rm dis}$ are the fermion-line connected and 
-disconnected quark Green functions, respectively. In Fig.~\ref{fig:green} we sketch both 
types of contributions. 
Note that (\ref{Gdisc}) only includes diagrams where gluon lines connect the quark loop
to the external legs. The unitary (full) quark Green function, including both connected 
and disconnected diagrams, is given by
\begin{equation}
G_{\mathcal{O}} = \left.\frac{\partial\,S(\lambda,\lambda)}{\partial\,\lambda}\right|_{\lambda=0} = 
G_{\mathcal{O}}^{\rm con} + G_{\mathcal{O}}^{\rm dis} \,.
\end{equation}

By multiplying $G_{\mathcal{O}}$ and $G_{\mathcal{O}}^{\rm con}$ with the inverse 
unmodified propagator from left and right we obtain singlet, 
\begin{equation}
\Gamma_{\mathcal{O}}^{\rm S} = S(0,0)^{-1} G_{\mathcal{O}}\,S(0,0)^{-1} \,,
\label{svert}
\end{equation}
and nonsinglet,
\begin{equation}
\Gamma_{\mathcal{O}}^{\rm NS} = S(0,0)^{-1} G_{\mathcal{O}}^{\rm con}\,S(0,0)^{-1} \,,
\label{nsvert}
\end{equation}
vertex functions. The corresponding renormalization factors are then given by 
\begin{equation}
{Z_{\mathcal{O}}^{\rm S}}^{-1}=\frac{1}{12}\,{\rm Tr}\,
\left[\Gamma_{\mathcal{O}}^{\rm S}\,{\Gamma_{\mathcal{O}}^{\rm Born}}^{-1}\right]\, Z_q^{-1} 
\label{zsin}
\end{equation}
and 
\begin{equation}
{Z_{\mathcal{O}}^{\rm NS}}^{-1}=\frac{1}{12}\,{\rm Tr}\,
\left[\Gamma_{\mathcal{O}}^{\rm NS} {\Gamma_{\mathcal{O}}^{\rm Born}}^{-1}\right]\, Z_q^{-1} \,.
\label{znonsin}
\end{equation}
We could have started from singlet and nonsinglet operators with a single parameter $\lambda$, 
as stated in (\ref{action}), instead of differentiating between operator insertions in propagator 
and determinant. For example
\begin{eqnarray}
\mathcal{O}^{\rm S}(x) &=& \sum_{q=u,d,s}\,\bar{q}(x)\,\Gamma\,q(x) \,, \label{sin}\\
\mathcal{O}^{\rm NS}(x) &=& \bar{u}(x)\,\Gamma\,u(x) - \bar{d}(x)\,\Gamma\,d(x) \,. \label{nonsin}
\end{eqnarray}
For the singlet operator (\ref{sin}) nothing changes. The nonsinglet operator (\ref{nonsin}) 
would contribute $O(\lambda^2)$ to the determinant for either choice of $M$, eqs.~(\ref{mdeg}) 
and (\ref{mmdeg}), which leaves us with
\begin{equation}
\left.\frac{\partial\,S(\lambda,\lambda)}{\partial\,\lambda}\right|_{\lambda=0} = G_{\mathcal{O}}^{\rm con} \,.
\end{equation}
We have just added the singlet operator to the action.
If we also added a term $\lambda^{\rm NS}_{\rm sea} \, O^{\rm NS}$ it would not change anything, the non-singlet operator
 would contribute to the determinant at $O((\lambda^{\rm NS}_{\rm sea})^2 )$, and so not change the
 derivative at $\lambda=0$.

\section{Numerical results and tests}

We shall now apply the Feynman-Hellmann method of nonperturbative renormalization 
to the axial vector current and the scalar density. It is convenient to introduce the primitive 
\begin{equation}
\Lambda_{\mathcal{O}}(\lambda_{\rm sea},\lambda_{\rm val}) =  
\frac{1}{12}\,{\rm Tr}\,\left[S(0,0)^{-1}\,S(\lambdas,\lambdav)\,S(0,0)^{-1}\,
{\Gamma_{\mathcal{O}}^{\rm Born}}^{-1}\right]\,.
\label{eq:Lam0}
\end{equation}
Expanding the propagator $S(\lambdas,\lambdav)$ in terms of $\lambdas, \lambdav$, using (\ref{exp}), we obtain  
\begin{equation}
\Lambda_{\mathcal{O}}(\lambdas,\lambdav) = a_0 + a_{\rm sea}\,\lambdas + 
a_{\rm val}\,\lambdav + O(\lambdas^2,\lambdas\,\lambdav,\lambdav^2)\,.
\label{eq:Lam1}
\end{equation}
The coefficients $a_{\rm sea}$ and $a_{\rm val}$ are what we need to compute,
\begin{equation}
Z_{\mathcal{O}}^{NS} = \frac{Z_q}{a_{\rm val}} \,, \quad Z_{\mathcal{O}}^{S} = \frac{Z_q}{a_{\rm val}+a_{\rm sea}} \,.
\label{eq:ZASNS}
\end{equation}
The proposed method involves the computation of two-point functions only. In the case of 
non\-singlet operators no extra gauge field configurations need to be generated. 
The parameters $\lambdas, \lambdav$ should be chosen large enough to give a strong signal, 
but small enough so that $\Lambda_{\mathcal{O}}$ can be fitted by a low-order polynomial in $\lambdas, \lambdav$. 

\begin{table}[!b]
  \begin{center}
    \renewcommand{\arraystretch}{1.5}
    \begin{tabular} {|c|c|c|c|c|}
       \hline
       $\lambdav$ & \multicolumn{4}{|c|}{$\lambdas$}  \\
       \hline
       $-0.0125\phantom{0000}$    &  $-0.03$ & $0.0$ & $0.00625$ & $0.0125$          \\[0.5ex]
       $-0.00625\phantom{000}$      & $-0.03$ & $0.0$ & $0.00625$ & $0.0125$          \\[0.5ex]
       $-0.003125\phantom{00}$ & $-0.03$ & $0.0$ & $0.00625$ & $0.0125$          \\[0.5ex]
       $\phantom{-}0.0\phantom{0000000}$  & $-0.03$ & $0.0$ & $0.00625$ & $0.0125$          \\[0.5ex]
       $\phantom{-}0.03\phantom{000000}$   & $-0.03$ & $0.0$ & $0.00625$ & $0.0125$          \\[0.5ex]
       \hline
    \end{tabular}
  \end{center}
  \caption{The parameters $\lambdav$ and $\lambdas$ employed in the simulations.}
  \label{lamTab}
\end{table}

The calculations are performed on $32^3\times 64$ lattices at $\beta=5.50$, corresponding to a lattice 
spacing of $a=0.074(2)\,\mbox{fm}$~\cite{Horsley:2013wqa}. We will use momentum sources~\cite{Gockeler:1998ye} 
throughout the calculation.
Using twisted boundary conditions, the momenta are chosen to be strictly diagonal, $p=(\rho,\rho,\rho,\rho)$. 
They are $(ap)^2=0.1542,\, 0.6169,\, 1.3879,\, 2.4674,\, 3.8553,\, 5.5517,\, 7.5564$ and $9.8696$, 
as given in the first column of Table III in~\cite{Constantinou:2014fka}. This choice of momenta 
will leave us with $O((ap)^2)$ scaling violations only, but with no direction-specific corrections, 
which we consider a great advantage. 

We are finally interested in renormalization factors in the RGI and $\overline{\rm MS}$ schemes. 
The conversion from the RI$^\prime$-MOM scheme to the RGI scheme is preferably done by a two-step process~\cite{Gockeler:2010yr}
\begin{equation}
Z_{\mathcal{O}}^{\rm RGI} = \Delta Z_{\mathcal{O}}^{\rm MOM}(\mu)\,
Z^{\rm MOM}_{\rm RI'-MOM}(\mu)\, Z_{\mathcal{O}}^{\rm RI'-MOM}(\mu)\,,
\label{conv1}
\end{equation}
which we follow here. The renormalization factors in the $\overline{\rm MS}$ scheme are given by
\begin{equation}
Z_{\mathcal{O}}^{\overline{\rm MS}}(\mu) = {\Delta Z^{\overline{\rm MS}}_{\mathcal{O}}(\mu)}^{-1}\, Z_{\mathcal{O}}^{\rm RGI} \,.
\label{conv2}
\end{equation}
The conversion factors $\Delta Z_{\mathcal{O}}^{\rm MOM}(\mu)$, $Z^{\rm MOM}_{\rm RI'-MOM}(\mu)$ and
$\Delta Z^{\overline{\rm MS}}_{\mathcal{O}}(\mu)$ are computed in continuum 
perturbation theory~\cite{Chetyrkin:1993hk,Larin:1993tq}. 
They depend on $\Lambda_{\overline{\rm MS}}$, which we choose as 
$\Lambda_{\overline{\rm MS}} = 339 \, {\rm MeV}$~\cite{Aoki:2013ldr}.

\begin{figure}[!hb]
\vspace*{0.25cm} \begin{center}
\includegraphics[scale=0.75,clip=true]{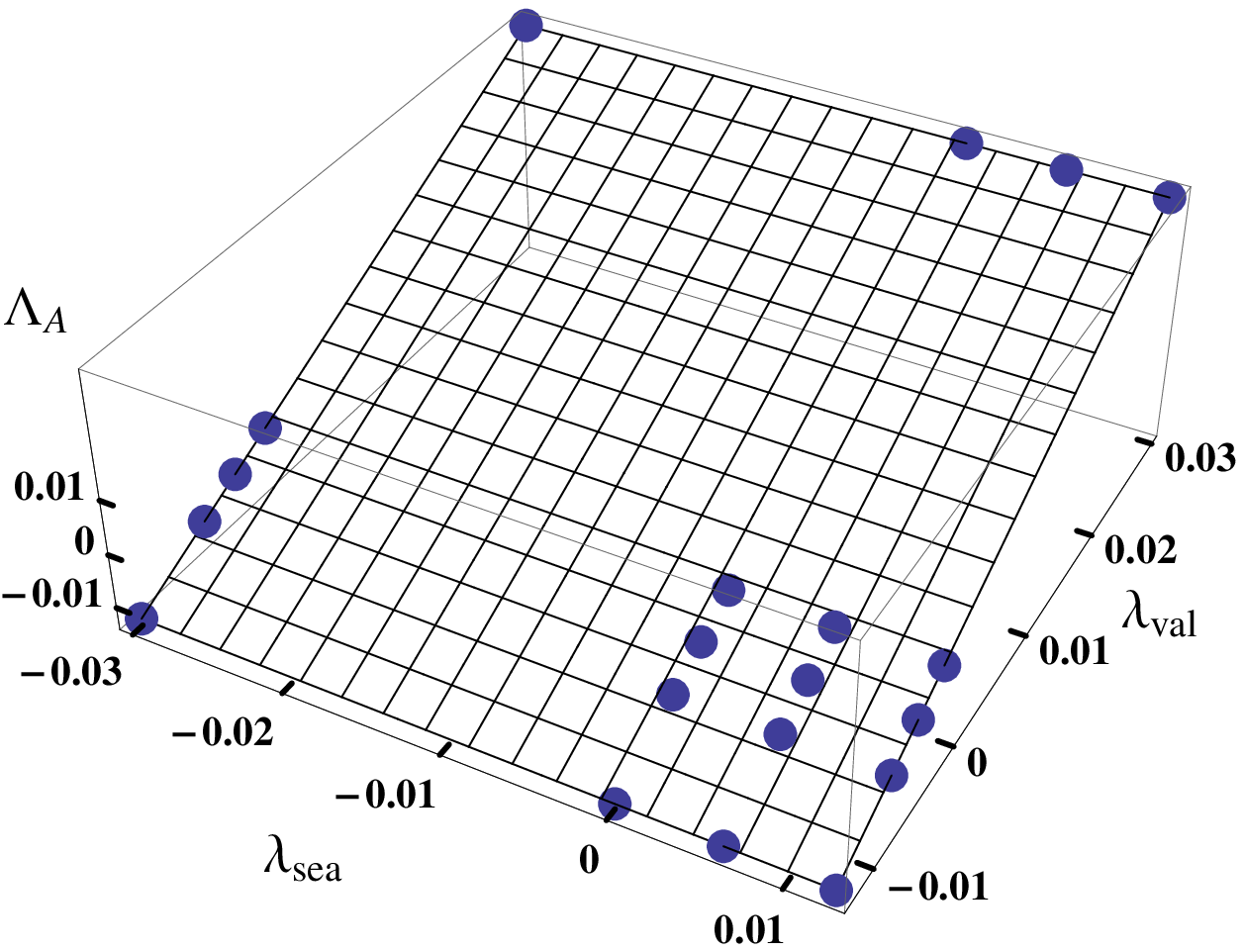}\\[2em]      
\includegraphics[scale=0.85,clip=true]{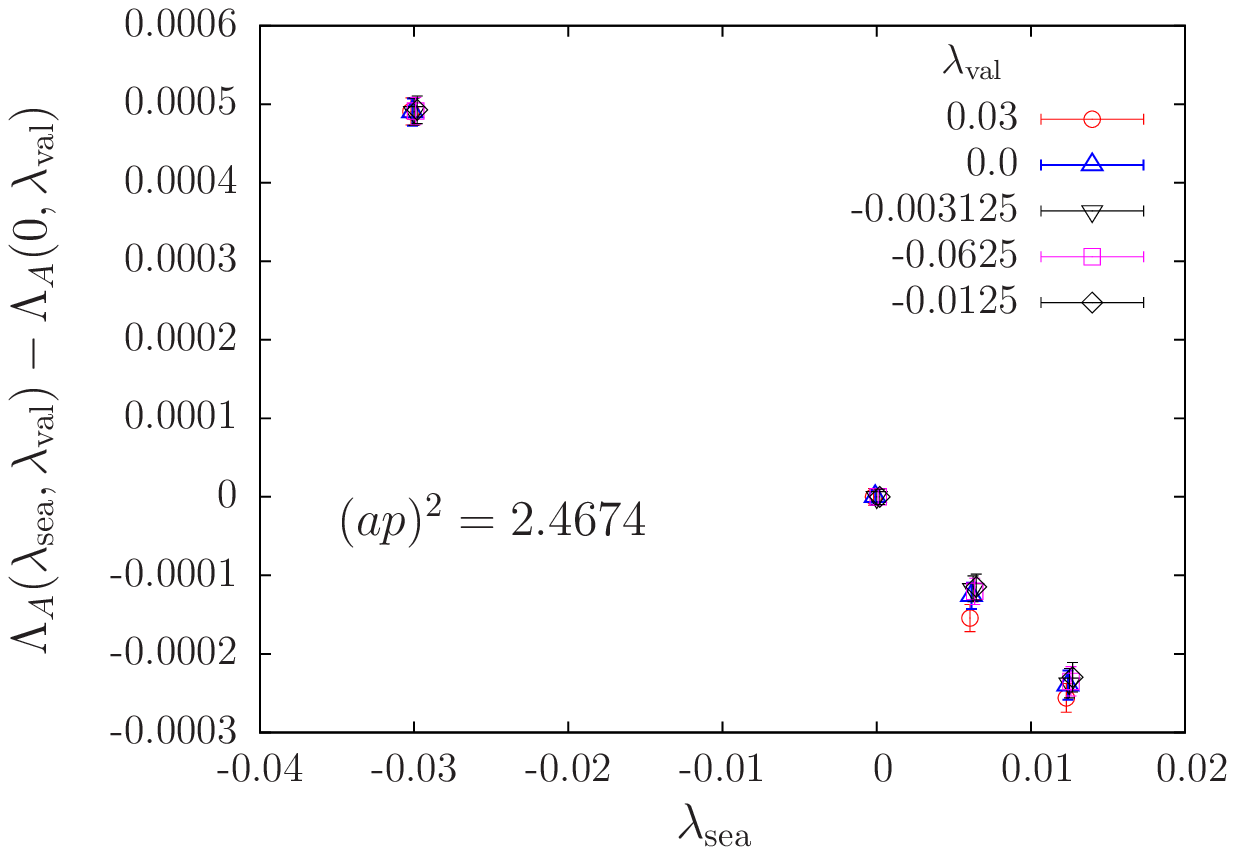}
\end{center}
\caption{Top panel: $\Lambda_{A}(\lambdas,\lambdav)$ as a function of 
$\lambdas$ and $\lambdav$ for $(ap)^2=2.4674$. Bottom panel: The difference 
$\Lambda_{A}(\lambdas,\lambdav)- \Lambda_{A}(0,\lambdav) = a_{\rm sea}\, \lambdas + O(\lambdas^2,\lambdas\lambdav,\lambdav^2)$ 
as a function of $\lambdas$, for $(ap)^2=2.4674$.}
  \label{fig:Lambda}
\end{figure}

\subsection{Axial vector current}
\label{sec:A3}

In order to proceed with the determination of the renormalization constant of the axial current, 
we add the third component of the axial current
\begin{equation}
   A_3(x) = \bar{q}(x)\, \gamma_3 \gamma_5\, q(x)\ ,
\end{equation}
to the action (\ref{action}). This operator is $\gamma_5$-hermitean, and hence suitable for inclusion as 
part of the Hybrid Monte Carlo when generating the new sets of gauge configurations required 
for the determination of the disconnected contributions. The simulations are performed at the SU(3) 
flavor symmetric point $\kappa_u=\kappa_d=\kappa_s=0.12090$~\cite{Bietenholz:2011qq}, corresponding 
to $m_\pi=m_K=465\,\mbox{MeV}$, for five different $\lambdav$ values with four different values of 
$\lambdas$ each. The actual run parameters are listed in Table~\ref{lamTab}.

\begin{figure}[!b]
  \begin{center}
     \begin{tabular}{cc}
        \includegraphics[scale=0.63,clip=true]{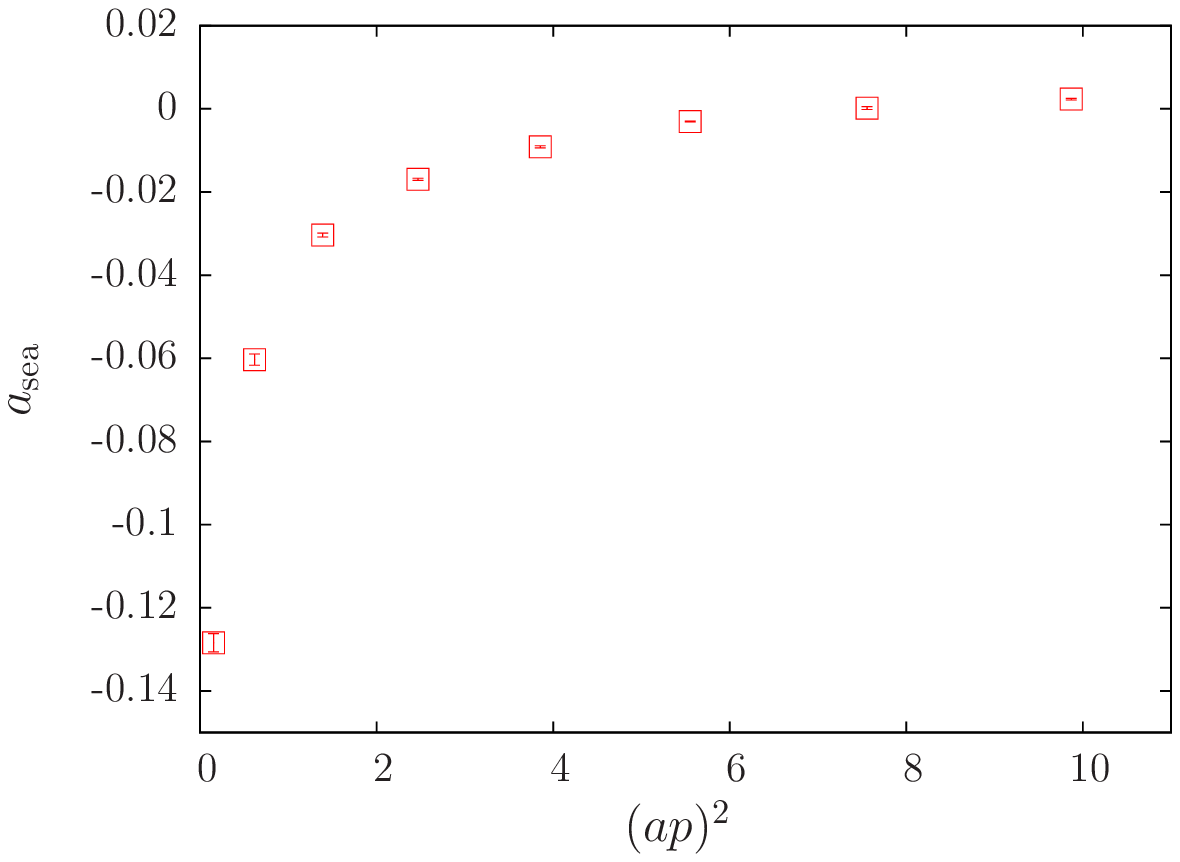}
        &
        \includegraphics[scale=0.63,clip=true]{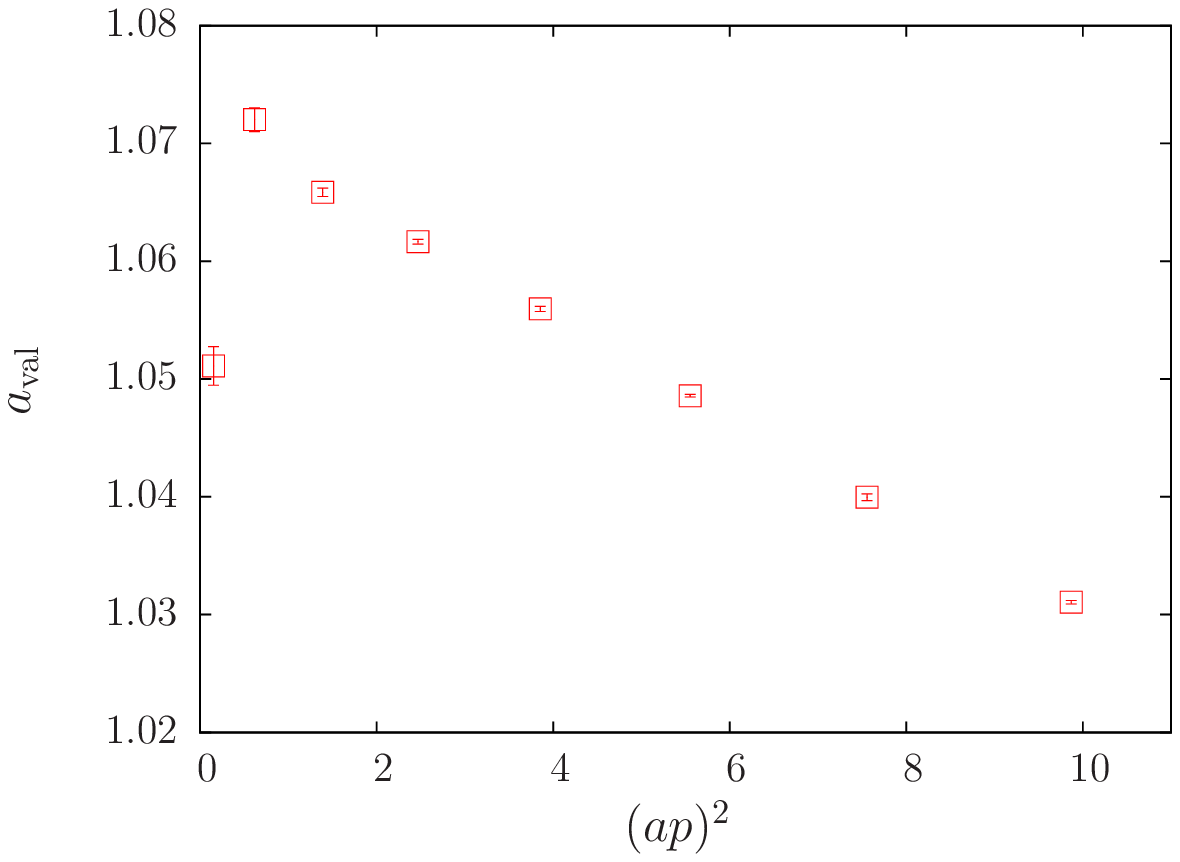}
     \end{tabular}
  \end{center}
  \caption{The coefficients $a_{\rm sea}$ and $a_{\rm val}$ as a function of $(ap)^2$.}
  \label{fig:asav}
\end{figure}

\begin{figure}[!t]
  \begin{center}
     \begin{tabular}{cc}
        \includegraphics[scale=0.63,clip=true]{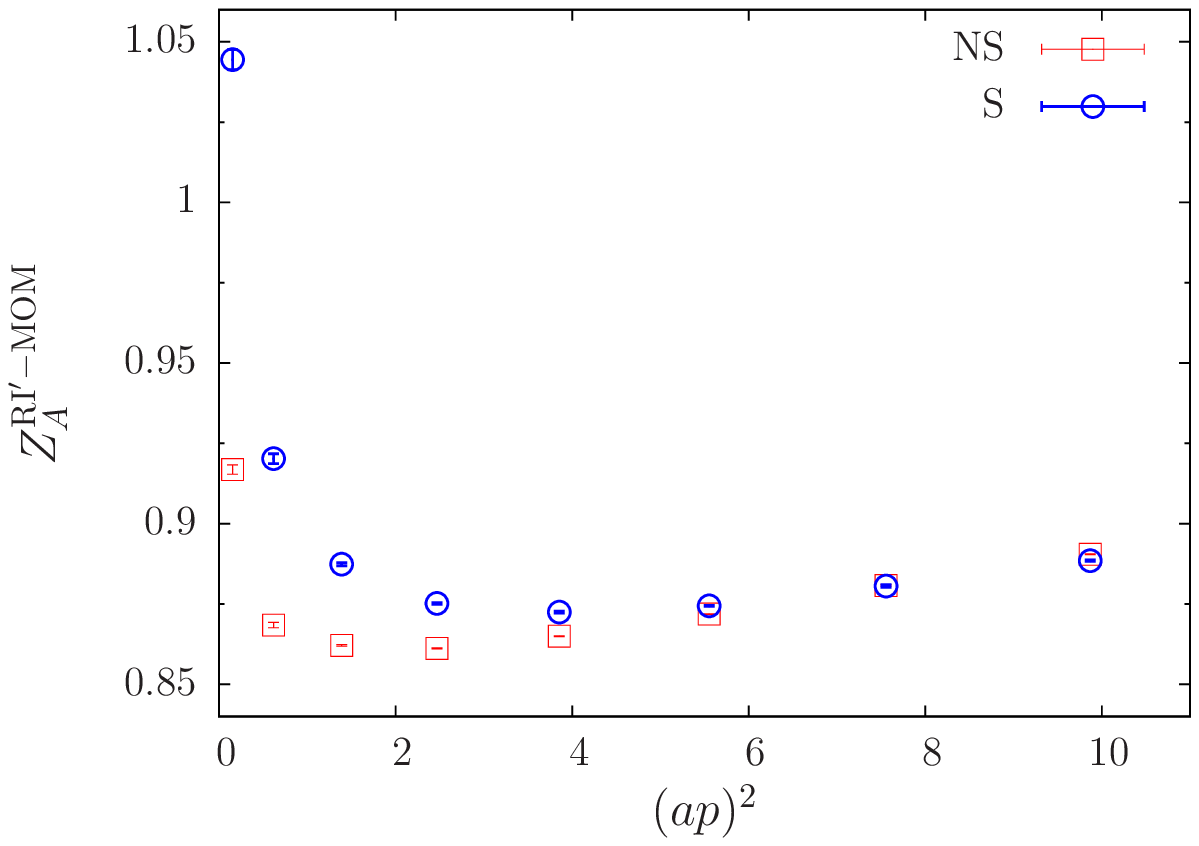}
        &
        \includegraphics[scale=0.63,clip=true]{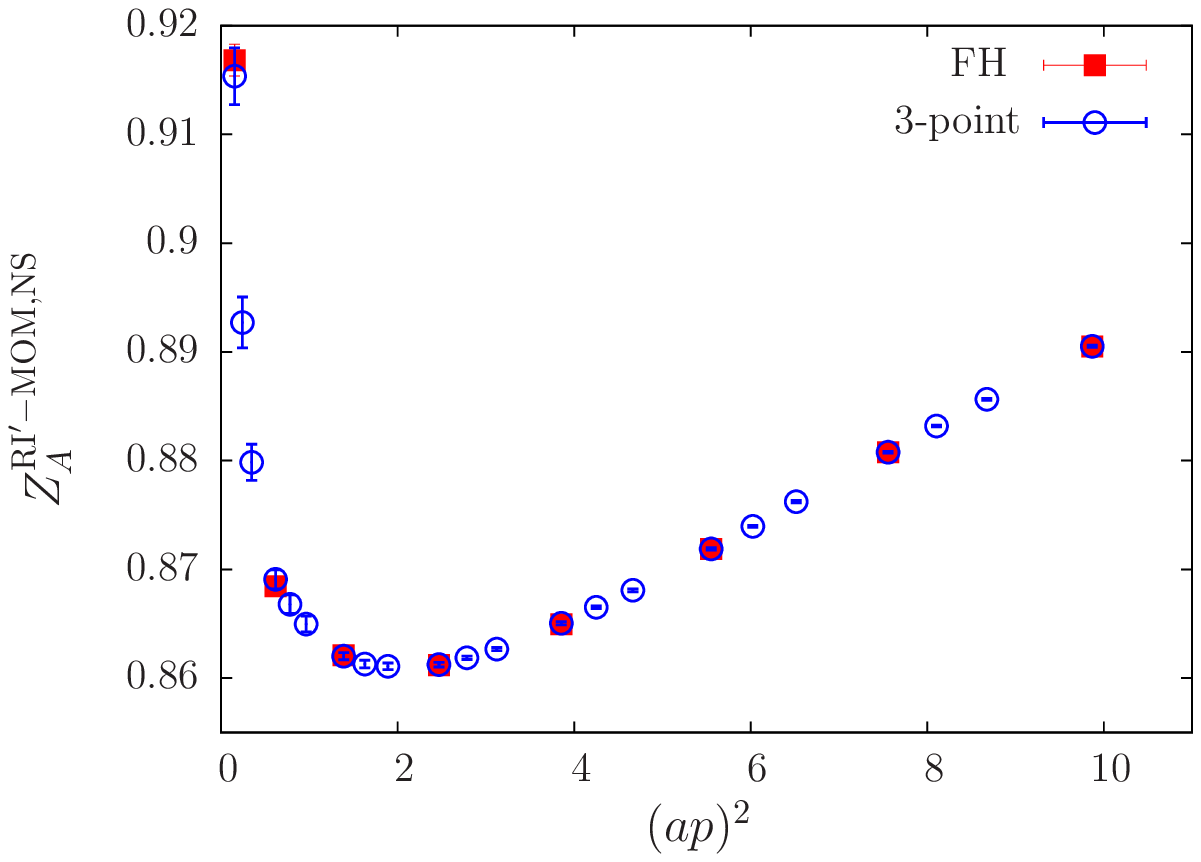}
     \end{tabular}
  \end{center}
\caption{Left panel: Singlet and nonsinglet renormalization factors $Z_A$ in the RI$^\prime$-MOM scheme 
at $\kappas=0.12090$. Right panel: Comparison of the nonsinglet renormalization factor $Z_A$ in 
the RI$^\prime$-MOM scheme obtained from the Feynman-Hellmann (FH) approach (this work) and the 
three-point function method~\cite{Constantinou:2014fka}.}
  \label{fig:ZAFH}
\end{figure}

In Fig.~\ref{fig:Lambda} we show our results for $\Lambda_A(\lambdas,\lambdav)$ and 
the difference $\Lambda_A(\lambdas,\lambdav) - \Lambda_A(0,\lambdav)$ for one of our intermediate momenta, 
$(ap)^2=2.4674$. Within the range of parameters we have explored, $\Lambda_A(\lambdas,\lambdav)$ 
(shown in the top figure) appears to be a linear function of both $\lambdas$ and $\lambdav$. 
The figure indicates that $a_{\rm sea} \ll a_{\rm val}$ for the axial vector current. 
In spite of being a rather small number, the disconnected contribution $a_{\rm sea}$ 
can be computed very accurately by our method. This is illustrated by the difference 
$\Lambda_A(\lambdas,\lambdav) - \Lambda_A(0,\lambdav) = a_{\rm sea}\, \lambdas + O(\lambdas^2,\lambdas\lambdav,\lambdav^2)$ 
(shown in the bottom figure). It helps the fit that higher order corrections are small. 
Similar results are found for the other momenta. We thus may fit our data for $\Lambda_A(\lambdas,\lambdav)$ 
by the ansatz
\begin{equation}
 \Lambda_A(\lambdav,\lambdas) = a_0 + a_{\rm sea}\, \lambdas + a_{\rm val}\, \lambdav \,.
   \label{eq:Lam2}
\end{equation}
This is done for each momentum source separately. The result is shown in Fig.~\ref{fig:asav}. From $a_{\rm sea}$ 
and $a_{\rm val}$, together with $Z_q$ defined in (\ref{eq:RIMOM2}), we obtain the renormalization factors in 
the RI$^\prime$-MOM scheme. The result is given in Fig.~\ref{fig:ZAFH} (left panel) for singlet and nonsinglet 
operators. The obvious question now is: how does that result compare with previous results using standard methods? 
In~\cite{Constantinou:2014fka} we have computed the nonsinglet renormalization factor from three-point functions 
using the same action. We compare that result with the Feynman-Hellmann result of this paper in 
Fig.~\ref{fig:ZAFH} (right panel). We find perfect agreement.

\begin{figure}[!b]
  \begin{center}
     \includegraphics[scale=0.85,clip=true]{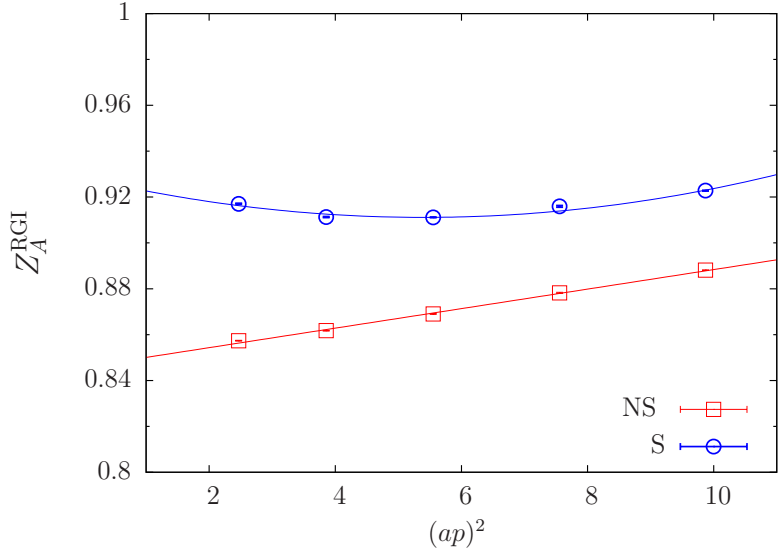}
  \end{center}
  \caption{Singlet and nonsinglet renormalization factors in the RGI scheme, together with a 
  linear (quadratic) fit to $2 \leq (ap)^2 \leq 10$ for the nonsinglet (singlet) $Z^{\rm RGI}$.}
  \label{fig:ZAFHRGI}
\end{figure}

Let us now convert our numbers to the RGI and $\overline{\rm MS}$ schemes, 
using (\ref{conv1}) and (\ref{conv2}). In the nonsinglet case 
$\Delta Z_A^{\rm MOM}(\mu) = \Delta Z_A^{\overline{\rm MS}}(\mu) =1$, 
as the anomalous dimension is zero. In the singlet case both 
$\Delta Z_A^{\rm MOM}(\mu)$ and $\Delta Z_A^{\overline{\rm MS}}(\mu)$ 
are nonzero and depend on the scale $\mu=\sqrt{p^2}$~\cite{Chetyrkin:1993hk,Larin:1993tq}. 
In Fig.~\ref{fig:ZAFHRGI} we show $Z_A^{\rm RGI}$ for both singlet and nonsinglet operators. 
We restrict ourselves to $(ap)^2 \geq 2$. Below that long-distance effects become dominant. 
As in~\cite{Constantinou:2014fka}, the nonsinglet data show scaling violations 
which can be approximated by a linear ansatz in $(ap)^2$.
We fit the singlet data by a quadratic ansatz. 
The result is 
\begin{equation}
  Z^{\rm RGI,\, NS}_A = 0.8458(8)\,, \quad
  Z^{\rm RGI,\, S}_A = 0.9285(36)\,.
\label{zrgi}
\end{equation} 
The renormalization factors $Z_A^{\overline{\rm MS}}(\mu)$ are obtained by multiplying the numbers 
in (\ref{zrgi}) by ${\Delta Z_A^{\overline{\rm MS}}(\mu)}^{-1}$. They are scale dependent. At $\mu=2\,\mbox{GeV}$ we obtain  
\begin{equation}
   Z^{\rm \overline{\rm MS},\, NS}_{A} = 0.8458(8) \,, \quad
   Z^{\rm \overline{\rm MS},\, S}_{A}=0.8662(34) \,.
\label{zms}
\end{equation}

The difference of singlet and nonsinglet renormalization factors of the axial vector current 
turns out to be small. That is not surprising since it is already known that in perturbation 
theory singlet and nonsinglet 
numbers start to depart only at two loops~\cite{Skouroupathis:2008mf}. The good news is that the 
Feynman-Hellmann method enables us to compute the disconnected contribution $a_{\rm sea}$, in spite of 
being a factor of $20$ smaller than the connected one $a_{\rm val}$, to an unprecedented precision of less than a percent.

It should be remembered that our results (\ref{zrgi}) and (\ref{zms}) refer to the flavor symmetric point
$\kappa_\ell = \kappa_s = 0.12090$.
To extrapolate the renormalization factors to the chiral limit, we would have to perform more simulations 
with the modified fermionic action at smaller quark masses.

\begin{table}[!t]
  \begin{center}
    \renewcommand{\arraystretch}{1.5}
    \begin{tabular} {|c|c|c|c|c|c|}
       \hline
       \multicolumn{5}{|c|}{$\kappa_{\rm val}$} & $\kappa_{\rm sea}$ \\
       \hline
       $0.120900$ & $0.120920$ & $0.120950$ & $0.120990$ & & $0.120900$ \\[0.5ex]
       & $0.190920$ & & & & $0.120920$  \\[0.5ex]
       & & $0.120950$ & & & $0.120950$  \\[0.5ex]
       & & & $0.120990$ & & $0.120990$  \\[0.5ex]
       & & & & $0.121021$ & $0.121021$  \\[0.5ex]
       \hline
    \end{tabular}
  \end{center}
  \caption{The parameters of background field configurations, $\kappa_{\rm val}$ and $\kappa_{\rm sea}$, used in the calculation
  of the scalar density.}
  \label{lam2Tab}
\end{table}

\subsection{Scalar density}
\label{sec:S}

We now turn to the scalar density
\begin{equation}
S(x) = \bar{q}(x) q(x) \,.
\label{sd}
\end{equation}
In this case the modification of the fermionic action, $S_F \rightarrow S_F - \lambda\,\sum_x S(x)$, is 
equivalent to changing the $\kappa$ values to $\kappa + \delta$, with $\delta =2\lambda\kappa^2/(1-2\lambda\kappa)$. 
As before, $\kappa_u=\kappa_d=\kappa_s$ is assumed. We allow the kappa values of sea and valence quarks to be different, 
and express the primitive (\ref{eq:Lam0}) in terms of the new variables $\deltas$ and $\deltav$. 
Expanding $\Lambda_{S}(\deltas,\deltav)$ about the reference point $(\kappa_{\rm sea},\kappa_{\rm val})$ then gives
\begin{equation}
\Lambda_{S}(\deltas,\deltav) = a_0 + \left(a_{\rm sea}/2\kappa_{\rm sea}^2\right)\,\deltas + 
\left(a_{\rm val}/2\kappa_{\rm val}^2\right)\,\deltav + O(\deltas^2,\deltas\,\deltav,\deltav^2)\,.
\label{eq:Lam3}
\end{equation}
Here we can draw on existing background gauge field configurations~\cite{Bietenholz:2011qq}. 
In Table~\ref{lam2Tab} we list the $\kappa$ parameters of the configurations used in this calculation.

\begin{figure}[!t]
  \begin{center}
     \includegraphics[scale=0.85,clip=true]{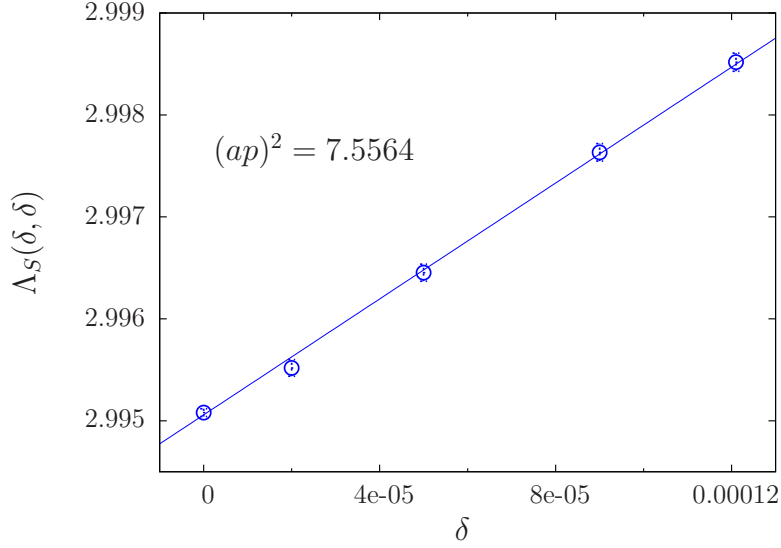}
  \end{center}
  \caption{The primitive $\Lambda_S(\delta,\delta)$ at the reference point 
  $\kappa_{\rm ref} =\kappa_{\rm sea}=\kappa_{\rm val}=0.12090$ as a function 
  of $\delta$ for  $(ap)^2=7.5564$, together with a linear fit.}
  \label{fig:ZSFH}
\end{figure}
\begin{figure}[!b]
  \begin{center}
       \includegraphics[scale=0.85,clip=true]{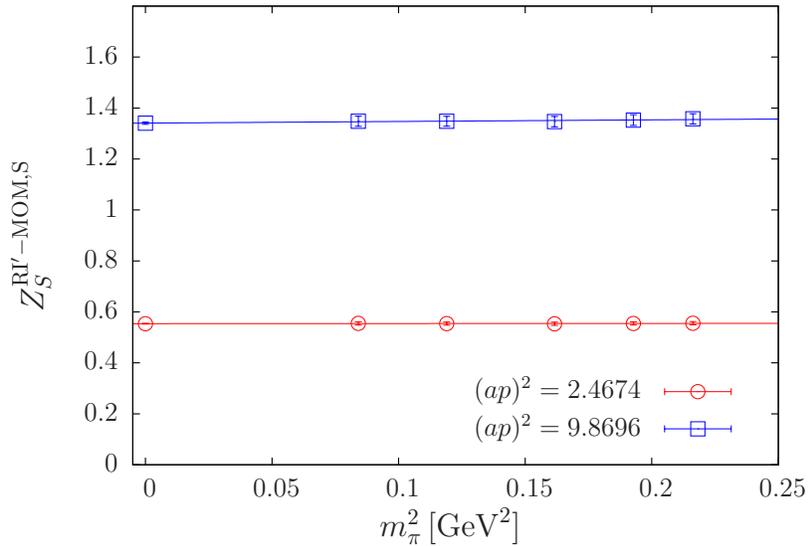}
  \end{center}
  \caption{The singlet renormalization factor in the RI$^\prime$-MOM scheme as
a function of $m_\pi^2$ for two momenta, $(ap)^2=2.4674$ and $9.869$, together with a linear extrapolation to the chiral limit.}
  \label{fig:ZSFH1}
\end{figure}

In Fig.~\ref{fig:ZSFH} we show $\Lambda_S(\delta,\delta)$ as a function of $\delta$ at the reference 
point $\kappa_{\rm ref} =\kappa_{\rm sea}=\kappa_{\rm val}=0.12090$ for one of our intermediate fit momenta, 
$(ap)^2=7.5564$. To a good approximation, the data lie on a straight line. 
From the slope at $\delta=0$ ($\kappa=\kappa_{\rm ref}$) we obtain the singlet renormalization factor 
in the  RI$^\prime$-MOM scheme,
\begin{equation}
\left.\frac{\partial\, \Lambda_S(\delta,\delta)}{\partial\,\delta}\right|_{\delta=0} = 
\frac{a_{\rm sea}+a_{\rm val}}{2\kappa_{\rm ref}^2} =  \frac{Z_q}{2\kappa_{\rm ref}^2\,Z_S^{\rm RI'-MOM,\, S}} \,.
\end{equation}
Repeating the calculation at $\kappa_{\rm ref}=0.12092, 0.12095, 0.12099$ and $0.121021$, with pion masses 
ranging from $465\,\mbox{MeV}$ ($\kappa=0.12090$) to $290\,\mbox{MeV}$ ($\kappa=0.121021$)~\cite{Constantinou:2014fka}, 
we can perform the chiral extrapolation of $Z_S^{\rm RI'-MOM,\, S}$. In Fig.~\ref{fig:ZSFH1} we show $Z_S^{\rm RI'-MOM, \, S}$ 
as a function of $m_\pi^2$ for two different momenta, together with the extrapolated values. 
Singlet $Z_S^{\rm RI'-MOM,\, S}$ is practically independent of the pion mass. 

\begin{figure}[!t]
  \begin{center}
        \includegraphics[scale=0.85,clip=true]{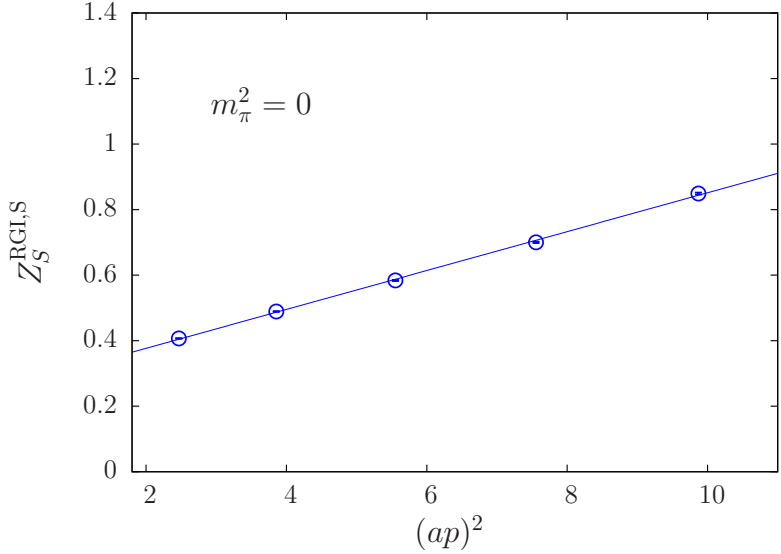}
  \end{center}
  \caption{The singlet renormalization factor $Z_S^{\rm RGI,\, S}$ in the chiral limit, 
  together with a linear fit to $2 \leq (ap)^2 \leq 10$.}
  \label{srgi}
\end{figure}

To convert $Z_S^{\rm RI'-MOM, \, S}$ to the RGI and $\overline{\rm MS}$ schemes we proceed as before. 
In Fig.~\ref{srgi} we show $Z_S^{\rm RGI,\, S}$. The data show scaling violations approximately linear in $(ap)^2$, 
which appear to be common to all our results~\cite{Constantinou:2014fka}. We restrict ourselves to $(ap)^2 \geq 2$ 
and fit the data by the ansatz $Z_S^{\rm RGI} + C\,(ap)^2$. The result is
\begin{equation}
Z_S^{\rm RGI,\, S} = 0.2617(35) \,,
\label{zsrgi}
\end{equation}
which upon conversion to the $\overline{\rm MS}$ scheme at $\mu = 2$ GeV gives
\begin{equation}
Z_S^{\rm \overline{\rm MS},\, S} = 0.3544(48)\,.
\label{zsms}
\end{equation} 
In contrast to (\ref{zrgi}) and (\ref{zms}), both numbers refer to the chiral limit.

As a further test, we have computed the nonsinglet renormalization factor 
$Z_S^{\rm RI'-MOM, \, NS}$ at $\kappa_{\rm ref}=0.12090$ and compared the outcome with our previous result 
from three-point functions~\cite{Constantinou:2014fka}. We find perfect agreement, as before. 

Using raw momentum data from~\cite{Constantinou:2014fka} we found 
in the chiral limit
\begin{equation}
Z_S^{\rm RGI,\, NS} = 0.5635(61)
\end{equation}
and
\begin{equation}
Z_S^{\rm \overline{\rm MS},\, NS} = 0.7631(82) \quad {\rm at} \quad \mu=2\, {\rm GeV}\,,
\end{equation}
giving
\begin{equation}
r_S=\frac{Z_S^{\rm RGI,\, NS}}{Z_S^{\rm RGI,\, S}} = 
\frac{Z_S^{\rm \overline{\rm MS},\, NS}}{Z_S^{\rm \overline{\rm MS},\, S}} = 2.15(4) \,.
\label{zsrat}
\end{equation}
Note that $\Delta Z_S^{\rm RGI}(\mu)=\Delta Z_S^{\overline{\rm MS}}(\mu)$. In continuum perturbation 
theory and for chiral fermions $r_S=1$. The deviation from one is an artifact of Wilson-type fermions. 
In~\cite{Gockeler:2004rp} it was found that $r_S$ rapidly approaches $r_S=1$ as the lattice spacing 
is decreased. An independent estimate of $r_S$ can be obtained from the ratio of valence to sea 
quark masses~\cite{Bietenholz:2011qq}. An updated value is $r_S=1.82(8)$, which is 
in reasonable agreement with the result (\ref{zsrat}).

\section{Conclusions}
\label{sec:con}

We have demonstrated that the Feynman-Hellmann method is an effective approach to calculating renormalization factors. 
For nonsinglet operators no additional gauge field configurations have to be generated. For singlet operators it appears 
that only a couple of different background field strengths need to be realized in order to make an accurate and precise 
calculation. We have demonstrated this through the determination of singlet and nonsinglet renormalization factors of 
the axial vector current and the scalar density.
Simulations of the axial vector current at smaller quark masses are in progress.

There is room for improvement. The renormalization factors show scaling violations  in $(ap)^2$, 
which has puzzled us already in~\cite{Constantinou:2014fka}. So far we have worked with unimproved quark propagators. 
Improving off-shell quark propagators should be simpler than improving three-point functions. 
Our goal is to remove lattice artifacts as far as possible. A first step in this direction has been 
taken in~\cite{Capitani:2000xi}.

\section*{Acknowledgements}

This work has been partly supported by the Deutsche Forschungsgemeinschaft, Grant SCHI 422/9-1, and 
the Australian Research Council, Grants FT100100005 and DP140103067. The numerical calculations were carried out on the
BlueGeneQ at NIC (J\"ulich, Germany), on the BlueGeneQ at EPCC (Edinburgh, UK) using DIRAC2 resources, 
and on the Cray XC30 at HLRN (Berlin and Hannover, Germany). Some of the simulations were undertaken on 
the NCI National Facility (Canberra, Australia), which is supported by the Australian Commonwealth Government.

\end{document}